\documentclass{iopart}
\usepackage{iopams}\usepackage{epsfig,graphicx}
    \newcommand{\ba}[1]{\begin{eqnarray}\label{#1}}

    \def\ee{\end{equation}}
    \def\ea{\end{eqnarray}}

\begin{document}

\title[Interdependence between integrable cosmological
models]{\textbf{Interdependence between integrable cosmological
models with minimal and non-minimal coupling}}

\author{Alexander~Yu~Kamenshchik$^{1,2}$, Ekaterina~O~Pozdeeva$^3$, Alessandro~Tronconi$^{1}$,
Giovanni~Venturi$^{1}$,  Sergey~Yu~Vernov$^{3}$}
\address{$^1$Dipartimento di Fisica e Astronomia and INFN, \small Via Irnerio 46, 40126, Bologna,
Italy,\\
$^2$L.D. Landau Institute for Theoretical Physics of the Russian
Academy of Sciences, Kosygin str. 2, 119334, Moscow, Russia,\\
$^3$Skobeltsyn Institute of Nuclear Physics,  Lomonosov Moscow State University, Leninskie Gory 1, 119991, Moscow, Russia}
\eads{\mailto{Alexander.Kamenshchik@bo.infn.it},
\mailto{pozdeeva@www-hep.sinp.msu.ru}, \mailto{Alessandro.Tronconi@bo.infn.it}, \mailto{Giovanni.Venturi@bo.infn.it}, \mailto{svernov@theory.sinp.msu.ru} }

\begin{abstract}
 We consider the relation between exact solutions of cosmological models having minimally and non-minimally coupled scalar fields. This is done for a particular class of solvable models which, in the Einstein frame, have potentials depending on hyperbolic functions and in the Jordan frame, where the non-minimal coupling is conformal, possess a relatively simple dynamics. We show that a particular model in this class can be generalized to the cases of closed and open Friedmann universes and still exhibits a simple dynamics. Further we illustrate the conditions for the existences of bounces in some sub-classes of the set of integrable models we have considered.
\end{abstract}



\submitto{\CQG}
\maketitle

\section{Introduction}

Recent observations~\cite{Planck2013,Planck2015} support interest in inflationary
models with non-minimal coupling between the inflaton field and the Ricci scalar curvature $R$. There are two  classes of non-minimally coupled inflationary models. In one of them a function of the scalar field (without a constant term) non-minimally coupled to $R$ is added to the Hilbert--Einstein term~\cite{HiggsInflation}, while in the other it replaces the latter~\cite{induced}.  This second class of models is known under the name of induced gravity and cosmology.

In spite of great success of numerical simulations and of using of different approximation schemes for studying of cosmological solutions of the Einstein equations, the exact analytical solutions of these equations are always useful. Indeed,  they permit to catch some important qualitative features of different cosmological models in the most easy and convenient way.

Most of the results of the exact integration of cosmological models with scalar fields are connected with minimally coupled models. A rather exhaustive list of such models was presented in paper \cite{Fre}.

The integrability of the models with minimal and non-minimal coupling was studied by using some advanced mathematical methods in \cite{Marek}.

In~\cite{KPTVV2013} we proposed a method for constructing integrable models with non-minimally coupled scalar fields by using the interrelation between the Jordan and Einstein frames. We considered some explicitly integrable cosmological models for spatially flat Friedmann universes filled with a minimally coupled scalar field, introduced in ~\cite{Fre}. On using the Einstein--Jordan frame transition, we constructed their counterparts in induced gravity models and in models having a conformally coupled scalar field plus the Hilbert--Einstein term.

In a recent paper \cite{Boisseau:2015hqa}, a flat Friedmann cosmological model in which the scalar field was conformally coupled to gravity and its potential included a positive cosmological term and a negative term with a quartic self-interaction was studied.  Such a model is exactly integrable and for a large class of initial conditions possesses a bounce and avoids
the cosmological singularity. Further, in such a model the cosmological evolution is, to a large extent, independent of the evolution of the scalar field. Let us note, that the presence of bounces was considered as a positive feature in cosmology and their study has rather a rich history~\cite{bounces}. In a subsequent paper \cite{Boisseau:2015cda}, the minimally coupled counterpart of the above model was considered. The corresponding potential is a linear combination of hyperbolic sine and cosine functions to the fourth power. It is curious that
the model with such a potential was previously studied in great detail in a series of papers   \cite{Bars:2010zh,Bars:2011th,Bars:2011mh,Bars:2012mt,Bars:2011aa}. In these papers such interesting topics as the possibility of the crossing of the Big Bang -- Big Crunch singularities and the existence of anti-gravity regimes were studied.
On the other hand one can see that this minimally coupled model belongs to a more wide class of models, described in the paper \cite{Fre} as a class 7. We did not consider the
non-minimally coupled counterpart of the models of this interesting class in our preceding paper \cite{KPTVV2013} and in the present paper we want to fill this gap.
Hence, we study the models of this class and find the conditions whereby they possesses a bounce.
Further, we show that the model, considered  in  \cite{Boisseau:2015hqa,Boisseau:2015cda} is easily generalizable to the cases of closed and open Friedmann universes, and present the corresponding solutions.    Remarkably, in this case the integrability of the non-minimally coupled model is more apparent that the integrability of its minimally coupled counterpart.

The structure of the paper is as follows: in Sect. 2 we review formulae that connect models in the Einstein and Jordan frames;
in Sect. 3 we describe the integrable cosmological model with a conformally coupled scalar field and its generalisation for the cases of closed and open Friedmann universes;
in Sect. 4 we discuss the properties of the cosmological solutions for a general class
 of minimally coupled models with potentials, depending on hyperbolic functions and for their non-minimally coupled counterparts;
 in Sect. 5 we study when the bounces are possible in these conformally coupled models;
 in Sect. 6 we write down some formulae the corresponding  induced gravity models.
The last Section contains concluding remarks.

\section{Relations between general exact solutions in the models with minimally coupled and non-minimally coupled scalar fields}

Let us consider a cosmological model, described by the following action
\begin{equation}
S =\int d^4x\sqrt{-g}\left[U(\sigma)R - \frac12g^{\mu\nu}\sigma_{,\mu}\sigma_{,\nu}-V(\sigma)\right],
\label{action}
\end{equation}
where $U(\sigma)$ and $V(\sigma)$ are differentiable functions of the scalar field $\sigma$.
If $U(\sigma)$ is a constant: $U(\sigma)=U_0$, then the scalar field is minimally coupled to gravity, otherwise we have a model with non-minimal coupling.

Let us consider the Friedmann--Lema\^{i}tre--Robertson--Walker (FLRW) metric. In this case the interval is
\begin{equation}
\label{Fried}
ds^2=N(\tau)^2 d\tau^2-a(\tau)^2\left(\frac{dr^2}{1-Kr^2}-r^2d\theta^2-r^2\sin^2(\theta)d\varphi^2\right),
\end{equation}
where $a(\tau)$ is the scale factor and $N(\tau)$ is the lapse function, $K$ is a constant. As usual $K=0$ describes a flat universe, $K=1$  a closed universe, and
$K=-1$ an open one.
On substituting this metric into the action (\ref{action}) and varying it with respect to $N$, $a$ and
$\sigma$, we obtain two Friedmann equations and a Klein-Gordon equation:
\begin{equation}
\label{Frequoc00}
6U\left[h^2+\frac{N^2}{a^2}K\right]+6hU'\dot{\sigma}=\frac12\dot{\sigma}^2+N^2V,
\end{equation}
\begin{equation}
\label{Frequocii}
\fl
 2U\left[2\dot{h}+3h^2-2h
 \frac{\dot{N}}{N}+\frac{N^2}{a^2}K\right]
+2U'\left[\ddot{\sigma}+2h\dot{\sigma}-\frac{\dot{N}}{N}\dot{\sigma}\right] =N^2V-\left[2U''+\frac12\right]
\dot{\sigma}^2,
\end{equation}
\begin{equation}
\ddot{\sigma}+\left(3h-\frac{\dot{N}}{N}\right)\dot{\sigma} -6U'\left[\dot{h}+2h^2-h\frac{\dot{N}}{N}+\frac{N^2}{a^2}K\right]+N^2V' = 0,
\label{KGoc}
\end{equation}
where $h=\dot{a}/a$, a ``dot'' means a derivative with respect to the parametric time $\tau$ and a ``prime'' means a derivative with respect to the scalar field.

If $U(\sigma)$ is not a constant, then the scalar field $\sigma$ is non-minimally coupled to gravity.
Let us perform a conformal transformation of the metric
\begin{equation}
g_{\mu\nu} = \frac{U_0}{U}\tilde{g}_{\mu\nu},
\label{conf}
\end{equation}
where $U_0$ is a positive constant.
We also introduce a new scalar field $\phi$ such that
\begin{equation}
\frac{d\phi}{d\sigma} = \frac{\sqrt{U_0(U+3U'^2)}}{U}
\quad\Rightarrow\quad
\phi = \int \frac{\sqrt{U_0(U+3U'^2)}}{U} d\sigma.
\label{scal1}
\end{equation}

As a result, the  action (\ref{action}) becomes the action for a minimally coupled scalar field:
\begin{equation}
S =\int d^4x\sqrt{-\tilde{g}}\left[U_0R(\tilde{g}) - \frac12\tilde{g}^{\mu\nu}\phi_{,\mu}\phi_{,\nu}-W(\phi)\right] ,
\label{action1}
\end{equation}
where
\begin{equation}
W(\phi) = \frac{U_0^2 V(\sigma(\phi))}{U^2(\sigma(\phi))}.
\label{poten}
\end{equation}
Let us emphasise that the formulae (\ref{conf})--(\ref{poten}) are valid for any metric.

In the Einstein frame the  FLRW metric (\ref{Fried}) becomes
\begin{equation}
\label{EinFreimMetric}
ds^2 ={}- \tilde{N}^2d\tau^2 + \tilde{a}^2\left(\frac{dr^2}{1-Kr^2}-r^2d\theta^2-r^2\sin^2(\theta)d\varphi^2\right),
\end{equation}
where the new lapse function and the new scale factor are defined as
\begin{equation}
\tilde{N} = \sqrt{\frac{U}{U_0}}N,\qquad \tilde{a} =  \sqrt{\frac{U}{U_0}}a.
\label{Na}
\end{equation}

For the case $K=0$ the Friedmann equations in the Einstein frame are:
\begin{equation}
6U_0\tilde{h}^2=\frac12\dot{\phi}^2+\tilde{N}^2W,
\label{Fried10}
\end{equation}
\begin{equation}
4U_0\dot{\tilde{h}}+6U_0\tilde{h}^2-4U_0\tilde{h}\frac{\dot{\tilde{N}}}{\tilde{N}}
= -\frac12\dot{\phi}^2+\tilde{N}^2W,
\label{Fried20}
\end{equation}
\begin{equation}
\ddot{\phi}+\left(3\tilde{h}-\frac{\dot{\tilde{N}}}{\tilde{N}}\right)\dot{\phi}
+\tilde{N}^2W'_{,\phi} = 0,
\label{KG0}
\end{equation}
where $\tilde{h} \equiv \dot{\tilde{a}}/{\tilde{a}}$.

Our goal is to obtain the general solutions for the Friedmann equations both in the Einstein and Jordan frames. We are assuming that such solutions are known in one frame and shall seek them in the other one. The algorithm has been proposed in our paper~\cite{KPTVV2013}. In this paper we consider new interesting examples and generalize our algorithm to the cases of open and closed Friedmann universes.

In~\cite{KPTVV2013} we considered in detail two specific forms of the function $U(\sigma)$.
For the induced gravity case~\cite{induced,we-ind-ex}
\begin{equation}
U_{ind}(\sigma) = \frac\gamma2  \sigma^2,
\label{induced}
\end{equation}
then we have from Eq. (\ref{scal1})
\begin{equation}
\fl
 \phi = \sqrt{\frac{2U_0(1+6\gamma)}{\gamma}}\ln \left[\frac{\sigma}{\sigma_0}\right]
\quad \mbox{and, \ inversely, } \quad
\sigma = \sigma_0 e^{\sqrt{\frac{\gamma}{2U_0(1+6\gamma)}}\phi}.
\label{connection_ind}
\end{equation}

We assume that $\gamma\neq -1/6$ since for the case of conformal coupling, $\gamma= -1/6$,  nontrivial solutions can exist only for the potential $V=V_0\sigma^4$~\cite{ABGV}.
 For $\gamma= -1/6$ we add a positive constant $U_0$ to the induced gravity term and consider~\cite{KPTVV2013}:
\begin{equation}
U_{c}(\sigma) = U_0 - \frac{\sigma^2}{12},
\label{conf-coupl}
\end{equation}
i.e. the case for which the coupling is conformal and a nonzero Einstein--Hilbert term is also present~\cite{Boisseau:2015hqa,we-ind-ex1,BNOS}.
For this case we get
\begin{equation}
\fl\phi = \sqrt{3U_0}\ln \left[\frac{\sqrt{12U_0}+\sigma}{\sqrt{12U_0}-\sigma}\right]
\quad\mbox{and }  \quad
\sigma = \sqrt{12U_0}\tanh\left[ \frac{\phi}{\sqrt{12U_0}}\right].
\label{connection_c}
\end{equation}

\section{An integrable model with a conformally coupled scalar field}

The model with $U(\sigma)=U_c$ and
the potential
\begin{equation}
\label{Vconf}
V_c=\Lambda-c\sigma^4\,,
\end{equation}
with $\Lambda>0$ and $c>0$ has been considered in~\cite{Boisseau:2015hqa}.
It was shown that the first Friedmann equation for this model reduces to the standard Friedmann equation where matter is represented by two perfect fluids: a positive cosmological constant and radiation, whose energy density can be both, positive or negative, depending on the initial conditions for the scalar field. If the energy density of the effective radiation is negative, then the universe undergoes a bounce.
In the subsequent paper \cite{Boisseau:2015cda}, it was shown that on applying a conformal transformation, combined with a reparametrisation of the scalar field (just as the procedure described in our paper \cite{KPTVV2013} and in the preceding section of the present paper), one arrives to a minimally coupled model with a potential
\begin{equation}
\label{Vint}
    W(\phi)  =\Lambda\cosh^{4}\left(
\frac{\phi}{2\sqrt{3U_0}}\right) - 144U_0^2c\sinh^{4}\left(
\frac{\phi}{2\sqrt{3U_0}}\right).
\end{equation}
 As we have already mentioned in the Introduction, this model was intensively investigated in papers \cite{Bars:2010zh,Bars:2011th,Bars:2011mh,Bars:2012mt,Bars:2011aa}. In particular, the corresponding inflationary scenario has been considered in~\cite{Bars:2010zh}.

Further, it  belongs to the set of integrable models~\cite{Fre} with a minimal coupling and a potential
\begin{equation}
W(\phi)=c_1\left(\cosh\left[\frac{3\beta\phi}{\sqrt{12U_0}}\right]\right)^{\frac{2(1-\beta)}
{\beta}}+c_2\left(\sinh\left[\frac{3\beta\phi}{\sqrt{12U_0}}\right]\right)^{\frac{2(1-\beta)}{\beta}},
\label{Wgeneral}
\end{equation}
where $c_1$,  and $c_2$ are arbitrary constants.
The model with the potential (\ref{Vint}) corresponds to the particular case of the potential
(\ref{Wgeneral}) with $\beta = 1/3$.
We shall discuss some properties of the above family of models in the subsequent sections.
Here we wish to stress that the main attractive feature of this class of models is their exact integrability in terms of the cosmic time parameter. Thus, it gives the opportunity of studying such fundamental questions of cosmology as the behavior of the universe in the vicinity of the cosmological singularity and the relations between different frames and different parameterizations of scalar fields.
The question whether the integrable one-field cosmological model classified in~\cite{Fre} can be embedded as consistent one-field truncations into Extended Gauged Supergravity or in $N = 1$ supergravity gauged by a superpotential has been considered in~\cite{Fre2}.

We now present the generalisation of the model of paper \cite{Boisseau:2015hqa} to the case of closed and open Friedmann universes.

The first Friedmann equation~(\ref{Frequoc00}) for this model is
\begin{equation}
\fl
6\left(U_0-\frac{1}{12}\sigma^2\right)\frac{\dot{a}^2}{a^2}-\frac{\dot{a}\dot{\sigma}\sigma}{a}+6\left(U_0-\frac{1}{12}\sigma^2\right)\frac{KN^2}{a^2}=\frac12\dot{\sigma}^2+N^2(\Lambda-c\sigma^4).
\label{Fried-closed}
\end{equation}
The Klein-Gordon equation~(\ref{KGoc}) is
\begin{equation}
\fl
\ddot{\sigma}+\left(3\frac{\dot{a}}{a}-\frac{\dot{N}}{N}\right)\dot{\sigma}+\sigma\left(\frac{\ddot{a}}{a}+\frac{\dot{a}^2}{a^2}\right)
-\sigma\frac{\dot{a}\dot{N}}{aN}-4cN^2\sigma^3+\frac{N^2K\sigma}{a^2}=0.
\label{KG-closed}
\end{equation}

On making a substitution such as the one done in paper \cite{Boisseau:2015hqa}: $\sigma={\chi}/{a}$, in the Klein-Gordon equation (\ref{KG-closed}), we reduce it to
\begin{equation}
\ddot{\chi}+\frac{\dot{\chi}\dot{a}}{a}-\frac{\dot{\chi}\dot{N}}{N}-\frac{4c\chi^3}{a^2}+\frac{N^2K\chi}{a^2}=0.
\label{KG-closed1}
\end{equation}
On choosing the lapse function $N$ as $N=a$, we can rewrite Eq. (\ref{KG-closed1}) in terms of the derivatives with respect to the conformal time $\eta$:
\begin{equation}
\frac{d^2\chi}{d\eta^2}-4c\chi^3+K\chi=0.
\label{ellip}
\end{equation}
On multiplying this equation by the conformal time derivative of $\chi$, $\frac{d\chi}{d\eta}$, we find its first integral
\begin{equation}
\frac12\left(\frac{d\chi}{d\eta}\right)^2-c\chi^4+\frac12K\chi^2 = A,
\label{ellip1}
\end{equation}
where $A$ is a constant.
Let us now rewrite the expression (\ref{ellip1}) in terms of the initial scalar field $\sigma$ and time derivatives in terms of the cosmic time (i.e. to the time parameter corresponding to the lapse function choice $N=1$). One obtains
\begin{equation}
 \frac12\dot{\sigma}^2+\frac12\sigma^2\dot{a}^2+\sigma\dot{\sigma}\frac{\dot{a}}{a}-c\sigma^4+\frac{K\sigma^2}{2a^2}=\frac{A}{a^4}.
 \label{ellip2}
 \end{equation}
On comparing Eq. (\ref{ellip2}) with the Friedmann equation  (\ref{Fried-closed}), we see that the latter reduces to a very simple form
 \begin{equation}
 6U_0\frac{\dot{a}^2}{a^2}+6U_0\frac{K}{a^2}=\Lambda+\frac{A}{a^4}.
 \label{Fried-closed1}
 \end{equation}
 This is a natural generalisation of the Friedmann equation, obtained in paper~\cite{Boisseau:2015hqa}. The equation for possible turning points (bounces or the points of maximal expansion) is
 \begin{equation}
 a^4-\frac{6U_0K}{\Lambda}a^2+\frac{A}{\Lambda}=0\,.
 \label{turn}
 \end{equation}
Let us give a list of the cosmological evolutions for different
 choices of the curvature $K$ and the radiation constant~$A$.\\
 1. (a). $K=0$, $A < 0$. We have a bounce at \\
 $$
 a_B=\left(\frac{-A}{\Lambda}\right)^{1/4}.
 $$
 This case was analysed in detail in papers \cite{Boisseau:2015hqa, Boisseau:2015cda}. \\
 (b) $K = 0$, $A=0$. For this case we have an infinite de Sitter expansion or an infinite  de Sitter contraction. The singularity is absent.\\
 (c) $K=0$, $A > 0$. We have an infinite expansion which begins from the Big Bang singularity or an infinite contraction which ends in the Big Crunch singularity.\\
2. (a) $K =1$, $A< 0$. We have a bounce at
$$
a_B=\left(\frac{3U_0}{\Lambda}+\sqrt{\frac{9U_0^2}{\Lambda^2}-\frac{A}{\Lambda}}\right)^{1/2}.
$$
(b) $K=1$, $A = 0$. We have a closed de Sitter universe which contracts,  has a bounce at
$$
a_B=\left(\frac{6U_0}{\Lambda}\right)^{1/2},
$$
and then expands infinitely.\\
(c) $K=1$, $0 < A < \frac{9U_0^2}{\Lambda}$. For this case we have two types of possible evolutions. One begins at the Big Bang singularity, expands until the point of maximal expansion
$$
a_M=\left(\frac{3U_0}{\Lambda}-\sqrt{\frac{9U_0^2}{\Lambda^2}-\frac{A}{\Lambda}}\right)^{1/2}
$$
and then contracts until the encounter with the Big Crunch singularity.  The second type of evolution is that with the bounce at
$$
a_B=\left(\frac{3U_0}{\Lambda}+\sqrt{\frac{9U_0^2}{\Lambda^2}-\frac{A}{\Lambda}}\right)^{1/2}.
$$
(d) $K=1$, $A = \frac{9U_0^2}{\Lambda}$. We have an Einstein static universe with radius
$$
a_S=\frac{3U_0}{\Lambda}.
$$
(e) $K = 1$, $A > \frac{9U_0^2}{\Lambda}$. We have an infinitely expanding or an infinitely contracting universe. \\
3. (a) $K = -1$, $A < 0$. We have a bounce at
$$
a_B=\left(-\frac{3U_0}{\Lambda}+\sqrt{\frac{9U_0^2}{\Lambda^2}-\frac{A}{\Lambda}}\right)^{1/2}.
$$
(b) $K=-1$, $A \geqslant 0$. We have an infinite expansion beginning from the Big Bang or an infinite contraction culminating in the Big Crunch.\\
For all these cases the Friedmann equation (\ref{Fried-closed1}) can be integrated explicitly, because the Ricci scalar as an integral of motion.
Combining Eqs.~(\ref{Frequoc00})--(\ref{KGoc}) it is easy to show that $R=R_0=2\Lambda/U_0$ at all values of $K$. We shall write down the explicit solution only for the case 2 (a). The expression for the scale factor is
\begin{equation}
a(t)=\left(\frac{3U_0}{\Lambda}+\sqrt{\frac{9U_0^2}{\Lambda^2}-\frac{A}{\Lambda}}\cosh \sqrt{\frac{\Lambda}{24U_0}}t\right)^{1/2}.
\label{Fried-closed2}
\end{equation}
 The Hubble parameter is
 \begin{equation}
 H(t)=\frac{\Lambda\sqrt{9U_0^2-A\Lambda}\sinh\left(\sqrt{\frac{\Lambda}{24U_0}}t\right)}
 {96U_0\left[3U_0+\sqrt{9U_0^2-A\Lambda}\cosh\left( \sqrt{\frac{\Lambda}{24U_0}}t\right)\right]}.
 \label{Fried-closed3}
 \end{equation}

Remarkably all the dependence of the cosmological evolution on the scalar field is encoded in the quantity $A$, i.e. it is nothing more than the evolution of the universe filled with the cosmological constant and the radiation-type fluid (see Eq.~(\ref{Fried-closed1})).

Let us notice that this ``disentanglement'' of the cosmological dynamics from the dynamics of the scalar field is a rather unusual phenomenon. For the case of models with a minimally coupled scalar field the only potential where this takes place is the constant potential, i.e. a cosmological constant. For the case of a conformally coupled scalar field the simple form of this Friedmann equation, which reduces to the standard one for a mixture of fluids, is connected with the fact that the conformally coupled scalar field behaves just like a radiation fluid with the constant $A$, which exhibits the properties of such a fluid. It is also important that quartic self-interaction term is also conformally invariant. The presence of this term gives us the opportunity of making the constant $A$ negative, by choosing a negative sign for the constant of the self-interaction of the scalar field.

\section{General solutions for a class of minimally coupled models with potentials, depending on hyperbolic functions and for their non-minimally coupled counterparts}

The integrability of the class of models with the potential (\ref{Wgeneral})  was discussed in~\cite{Fre}.
We shall arrive to this result in a slightly different way and also write down the corresponding formulae for the conformally coupled counterpart of this class.


The minisuperspace Lagrangian, generating the Friedmann equations  (\ref{Fried10}), (\ref{Fried20}) and the Klein-Gordon equation (\ref{KG0}) is
\begin{equation}
L=\frac{6\left(\dot{\tilde{a}}\right)^2\tilde{a}U_0}{\tilde{N}}
-\frac{\tilde{a}^3\dot{\phi}^2}{2\tilde{N}}+\tilde{N}W\tilde{a}^3.
\end{equation}

If one considers the potential (\ref{Wgeneral}) and chooses the lapse functions
\begin{equation*}
\tilde{N}=\frac{4U_0}{3\beta^2}\tilde{a}^{3-6\beta},
\end{equation*}
then the Lagrangian has the following form:
\begin{equation}\label{LL}
L=\frac92\beta^2\dot{\tilde{a}}^2\tilde{a}^{6\beta-2}
-\frac{3\beta^2\tilde{a}^{6\beta}\dot{\phi}^2}{8U_0}+\frac{4U_0}{3\beta^2}W\tilde{a}^{6-6\beta}.
\end{equation}

Let us introduce new variables $x$ and $y$ defined as
\begin{equation}
\tilde{a}^{6\beta}=xy,\qquad \exp\left(\frac{6\beta}{\sqrt{12U_0}}\phi\right)=\frac{x}{y},
\end{equation}
then the Lagrangian (\ref{LL}) takes the form
\begin{equation}
\label{LagXY}
L=\frac12\dot{x}\dot{y}+\frac{4U_0}{3\beta^2} (xy)^{\frac{1-\beta}{\beta}}W\,.
\end{equation}

For the potential $W$, given by (\ref{Wgeneral}),
we finally obtain, on introducing another couple of independent variables
\begin{equation}
 \xi=\frac{x+y}{2},\quad \eta=\frac{x-y}{2}\, ,
 \end{equation}
the following simple expression for the Lagrangian:
\begin{equation}
L=\frac{\dot{\xi}^2-\dot{\eta}^2}{2}+\frac{4U_0}{3\beta^2}\left(c_1\xi^{\frac{2(1-\beta)}{\beta}}
+c_2\eta^{\frac{2(1-\beta)}{\beta}}\right).\label{Lagrangian}
\end{equation}

The corresponding Euler--Lagrange equations are
\begin{equation}
\label{equxieta}
\fl\ddot{\xi}-c_1\frac{4U_0}{3\beta^2}\left(\frac{2(1-\beta)}{\beta}\right)\xi^{\frac{2}{\beta}-3}=0,\qquad
\ddot{\eta}+c_2\frac{4U_0}{3\beta^2}\left(\frac{2(1-\beta)}{\beta}\right)\eta^{\frac{2}{\beta}-3}=0.
\end{equation}

Their first integrals are
\begin{equation}
\frac{\dot{\xi}^2}{2}-c_1\frac{4U_0}{3\beta^2}\xi^{\frac{2(1-\beta)}{\beta}}=E_1,\qquad\frac{\dot{\eta}^2}{2}+c_2\frac{4U_0}{3\beta^2}\eta^{\frac{2(1-\beta)}{\beta}}=E_2.
\end{equation}

These equations have solutions by quadrature. For $\beta=1/3$ one obtains the solutions in terms of elliptic functions~\cite{Bars:2010zh}.

The case $\beta=1/2$ is also well-known. For this case the parametric time coincides with the cosmic time and the solutions can be expressed in terms of elementary functions
(see e.g. \cite{Giampiero} and references therein).

It immediately follows from the first Friedmann equation that $E_1=E_2$.
We can now find the cosmological variables
\begin{equation*}
\fl\tilde{a}=\left(\xi^2-\eta^2\right)^{1/(6\beta)},\quad  \tilde{N}=\frac{4U_0}{3\beta^2}\left(\xi^2-\eta^2\right)^{(1-2\beta)/(2\beta)},\quad \phi=\frac{\sqrt{3U_0}}{3\beta}\ln\left(\frac{\xi+\eta}{\xi-\eta}\right).
\end{equation*}

One can also write down the expressions for the solutions of the corresponding models with non-minimal coupling. Indeed, on using (\ref{Na}) and (\ref{connection_c}), we obtain
\begin{eqnarray}
  a_c&=&\frac{1}{2}\left((\xi+\eta)^{1/(3\beta)}+(\xi-\eta)^{1/(3\beta)}\right), \\
 N_c&=&\frac{2U_0}{3\beta^2}\left((\xi+\eta)^{1/(3\beta)}+(\xi-\eta)^{1/(3\beta)}\right)
 \left(\xi^2-\eta^2\right)^{(1-3\beta)/(3\beta)}
,\\
 \sigma_c&=&\sqrt{12U_0}\frac{(\xi+\eta)^{1/(3\beta)}-(\xi-\eta)^{1/(3\beta)}}
{(\xi+\eta)^{1/(3\beta)}+(\xi-\eta)^{1/(3\beta)}}\, ,
\end{eqnarray}
for the model with conformal coupling. At $\beta=1/3$ we get $a_c=\xi$.

The associated explicit expressions for the potential $V_c$ is
\begin{equation}
\fl\eqalign{V_c(\sigma)=\frac{1}{36U_0^2 4^{1/\beta}}\left\{c_1\frac{\left[(\sqrt{12U_0}+\sigma)^{3\beta}
+(\sqrt{12U_0}-\sigma)^{3\beta}\right]^{\frac{2(1-\beta)}{\beta}}}{(12U_0-\sigma^2)^{1-3\beta}}+{}\right.\\\left.\qquad {} +c_2\frac{\left[(\sqrt{12U_0}+\sigma)^{3\beta}
-(\sqrt{12U_0}-\sigma)^{3\beta}\right]^{\frac{2(1-\beta)}{\beta}}}{(12U_0-\sigma^2)^{1-3\beta}}\right\}.}
\label{V-c}
\end{equation}
and corresponds to the potential (\ref{Wgeneral}).
In the next section we shall consider some general properties of such potentials.

\section{Integrable models with bounce solutions}
Let us consider the model with a conformal coupling, the Hilbert--Einstein term and the potential
$V_c$, given by (\ref{V-c}).
We wish to find the conditions for the existence of bounces in such models. Let us first consider the Friedmann equations and the Klein-Gordon equation for an arbitrary model, wherein the scalar field is conformally coupled to the scalar curvature.

A bounce point $t_B$ is defined by two conditions: the Hubble parameter $H=0$ and $\dot{H}>0$.
On substituting $N=1$ and $H=h=0$ into  system (\ref{Frequoc00})--(\ref{KGoc})  with $K=0$, we get
\begin{equation}
\frac12\dot{\sigma}^2+V=0,
\label{bounce}
\end{equation}
\begin{equation}
4U\dot{H}+\left(2U''+\frac12\right)\dot{\sigma}^2+2U'\ddot{\sigma}-V=0,
\label{bounce1}
\end{equation}
\begin{equation}
\ddot{\sigma}-6U'\dot{H}+V'=0.
\label{bounce2}
\end{equation}

On substituting $U=U_c$ and using (\ref{bounce}) and (\ref{bounce2}),  we obtain from Eq. (\ref{bounce1})
\begin{equation}
4U_0\dot{H}-\frac43V+\frac13\sigma V'=0.
\label{bounce4}
\end{equation}
From Eq. (\ref{bounce4}) one has
\begin{equation}
\dot{H}=\frac{1}{12U_0}\left(4V-\sigma V'\right).
\label{bounce5}
\end{equation}
 Thus,
we have the condition
\begin{equation}
4V-\sigma V' > 0.
\label{bounce6}
\end{equation}
Further, it follows from Eq. (\ref{bounce}) that at the point where $H=0$, the potential
$V<0$.

We can now consider some particular values of the parameter $\beta$, for which the potential (\ref{V-c}) appears relatively simple. \\
\textbf{For the case $\beta=1/3$,} the potential (\ref{V-c}) can be written as
\begin{equation}
V=V_1+V_2\sigma^4.
\label{1-3}
\end{equation}
On substituting formula (\ref{1-3}) into formula (\ref{bounce6}) we obtain the condition
$$
V_1>0.
$$
Further, in order to have a negative potential (\ref{1-3}) at some value of the scalar field it is necessary to have $V_2 < 0$.
This is just the case considered in \cite{Boisseau:2015hqa, Boisseau:2015cda}.\\
\textbf{For the case $\beta = 1$,} the potential (\ref{V-c}) can be written as
\begin{equation}
V=V_0\left(U_0-\frac{\sigma^2}{12}\right)^2,
\label{1}
\end{equation}
where $V_0=(c_1+c_2)/U_0^2$.
This potential can be negative if and only if the constant $V_0<0$. Then, on substituting the potential (\ref{1}) into the condition (\ref{bounce6}), we obtain
\begin{equation}
12U_0<\sigma^2.
\label{pot-12}
\end{equation}

\textbf{Another interesting value}  is $\beta = 2/3$.
In this case the potential can be written as
\begin{equation}
V=\frac{c_1}{144U_0^2}(12U_0-\sigma^2)\left(\sigma^2+12U_0+2V_1\sigma\right),
\label{pot-23}
\end{equation}
where $V_1=2c_2\sqrt{3U_0}/c_1$. The study of the conditions for the existence of bounces with arbitrary values of the constants $c_1$  and $V_1$ is rather cumbersome. Thus, we shall limit ourselves to considering  a few particular values of the constant $V_1$.

Firstly, let us consider the case, when
\begin{equation*}
V_1={}-\sqrt{12U_0}.
\end{equation*}

Now, the potential $V$ can be written down as
\begin{equation*}
V=\frac{c_1}{144U_0^2}\left(12U_0-\sigma^2\right)\left(\sigma-\sqrt{12U_0}\right)^2,
 \label{pot-230}
 \end{equation*}
 while
 \begin{equation*}
 4V-\sigma V' = 2\frac{\sqrt{12U_0}c_1}{144U_0^2} \left(\sigma-\sqrt{12U_0}\right)^2\left(\sigma+2\sqrt{12U_0}\right).
 \end{equation*}

 In this case  two conditions of the existence of bounce are compatible if
  \begin{equation*}
 c_1 > 0,\qquad \sigma>\sqrt{12U_0}\ \quad {\rm or}\ \quad -2\sqrt{12U_0} < \sigma < -\sqrt{12U_0}.
 \end{equation*}

 Secondly, let us consider the case
  \begin{equation*}
 V_1=\sqrt{12U_0}.
 \end{equation*}
 Now the potential is
 \begin{equation*}
 V=\frac{c_1}{144U_0^2}\left(12U_0-\sigma^2\right)\left(\sigma+\sqrt{12U_0}\right)^2
 \end{equation*}
 and
  \begin{equation*}
 4V-\sigma V' =
 {}-2\sqrt{12U_0} \frac{c_1}{144U_0^2}\left(\sigma-2\sqrt{12U_0}\right)\left(\sigma+\sqrt{12U_0}\right)^2.
 \end{equation*}
 The conditions for the existence of bounces are compatible if
 \begin{equation}
 c_1 > 0,\qquad \sqrt{12U_0} < \sigma < 2\sqrt{12U_0}.
 \end{equation}

\begin{figure}[h] \centering
\includegraphics[height=5.27cm]{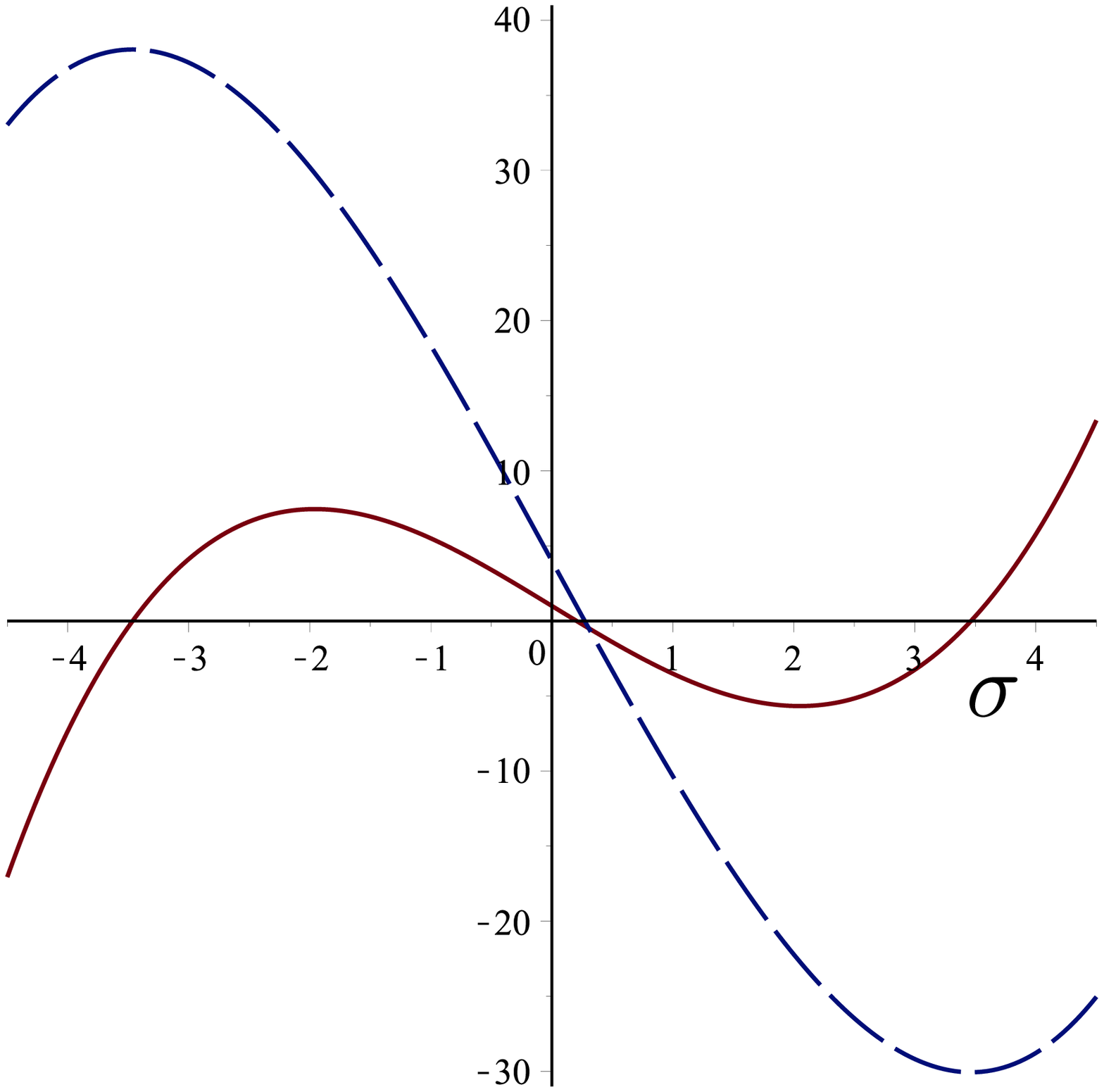}\  \ \includegraphics[height=5.27cm]{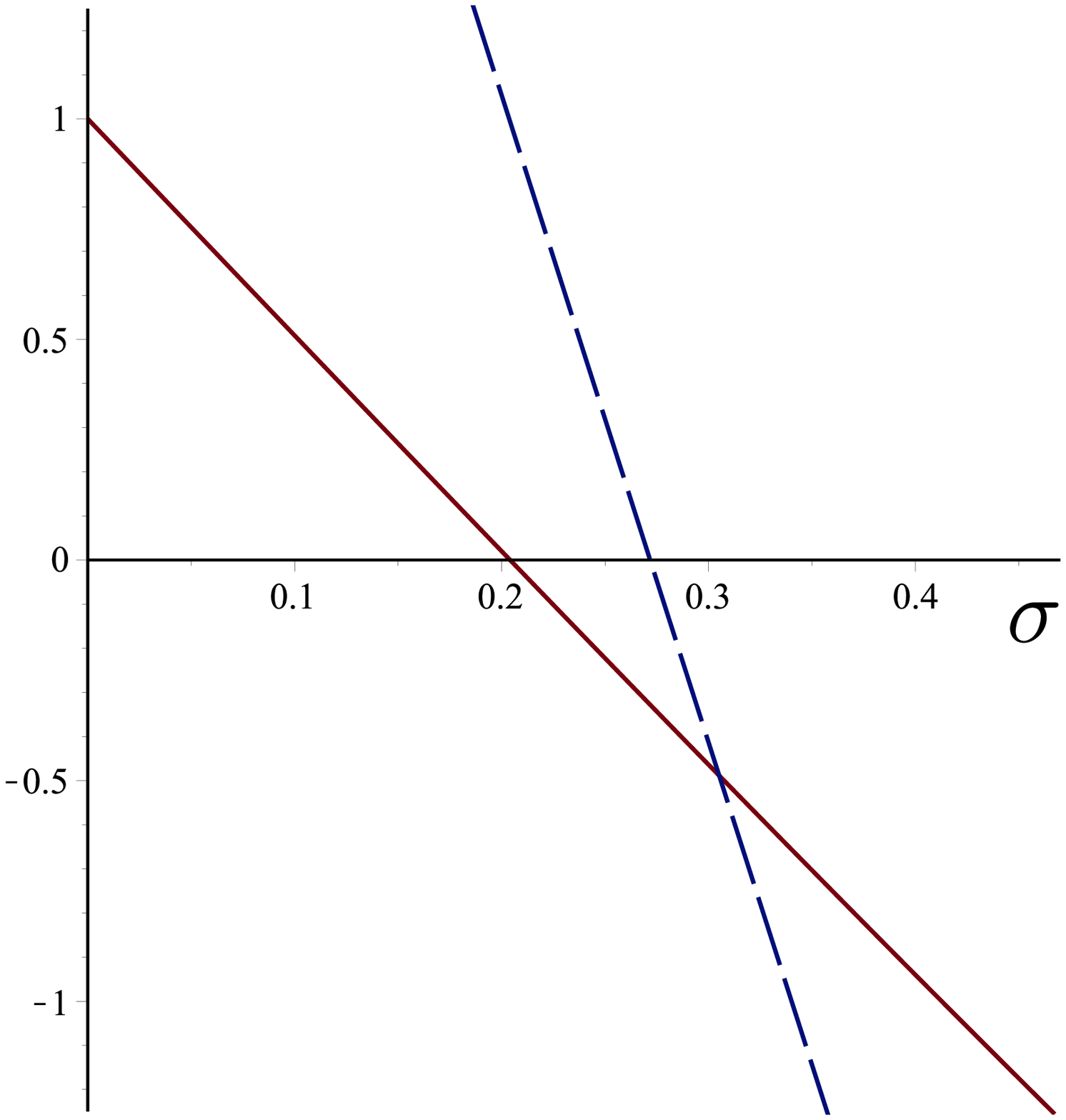}
\caption{ The potential $V_c$ (solid red line) and the expression $4V_c-\sigma V^\prime_c$ (dash blue line) at $U_0=1$, $c_1=1$, $\beta=2/3$,
$V_1=-289\sqrt{6}/24$. The fine structure, presented on the right picture, shows that a bounce is possible.}
\label{Fig23}
\end{figure}

We see that in these cases as well as for $\beta=1$ the bounce corresponds to a such $\sigma$ that $U<0$.
The following example shows that for $U>0$ the bounce is possible not only in the case $\beta=1/3$.
Let
\begin{equation}
c_2 = {}-\frac{289\sqrt{2}}{48}c_1,\qquad \Rightarrow\qquad V_1=-\frac{289}{24}\sqrt{6U_0}.
\end{equation}
For  positive values of $\sigma$ we get that $V<0$ at
\begin{equation}
\frac{1}{12}\sqrt{6U_0}<\sigma<\sqrt{12U_0}.
\end{equation}
At $U_0=1$,  $\frac{1}{12}\sqrt{6U_0}\simeq 0.20412$.
At the same time the condition (\ref{bounce6}) is satisfied if $\sigma<0.27178$ (see Fig.~\ref{Fig23}).
Thus, we get a bounce point at $U>0$.

\begin{figure}[h] \centering
\includegraphics[height=5.27cm]{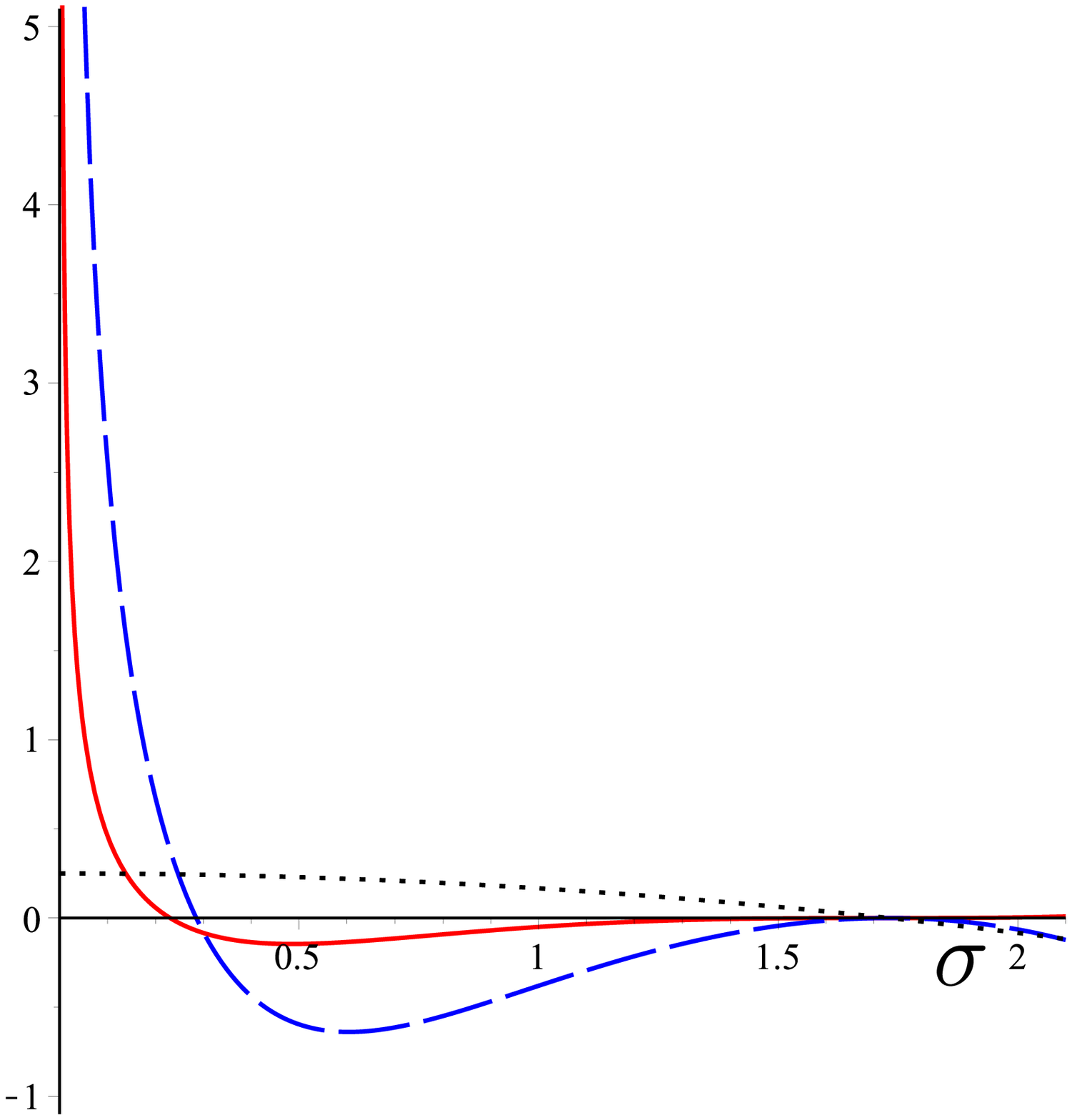}\ \  \includegraphics[height=5.27cm]{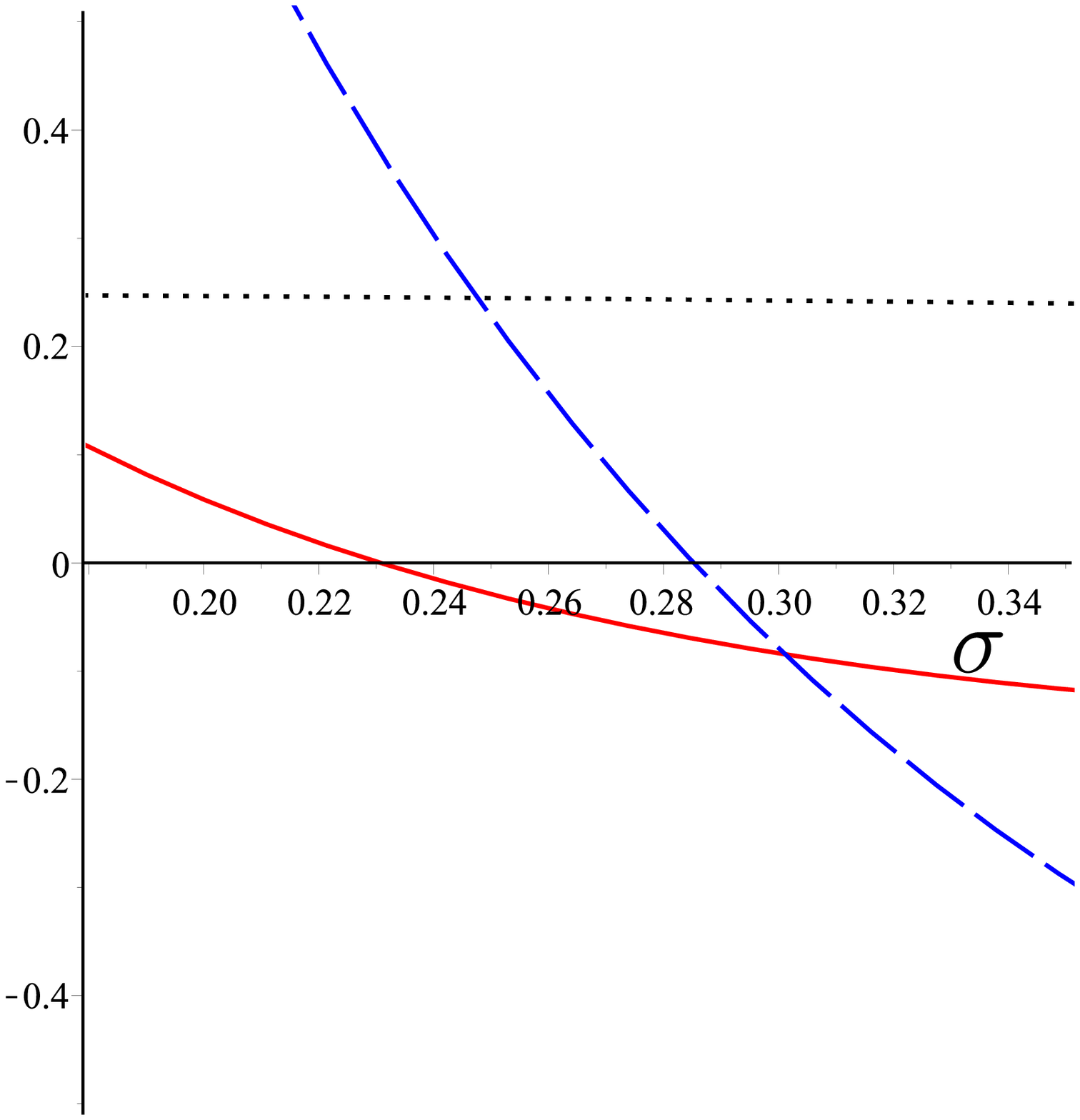}
\caption{ The potential $V_c$ (solid red line) and the expression $4V_c-\sigma V^\prime_c$ (dash blue line) and the function $U$ (dot black line) at $U_0=1/4$, $c_1=-1$, $c_2=0.7$, and $\beta=4/3$. The fine structure, presented on the right picture, shows that a bounce is possible.}
\label{Fig43}
\end{figure}

\textbf{For the case} $\beta = 4/3$ the potential can be represented as:
\begin{equation*}
\fl V_c={\frac {\sqrt {2} \left(12U_0-{\sigma}^{2} \right) ^{3} }{576{U_0}^{2}}}\,\left[
{\frac {{c_2}}{\sqrt {12\,\sqrt {3}{U_0}^{3/2}\sigma+\sqrt {3}
\sqrt {U_0}{\sigma}^{3}}}}+{\frac{4{c_1}}{\sqrt {2\,{\sigma}^{4}+144
\,U_0{\sigma}^{2}+288\,{U_0}^{2}}}} \right]
\end{equation*}
In this case it is difficult to obtain solutions in terms of elementary functions. We consider
it by using numerical calculations. For $U_0=1/4$, $c_1=-1$, and $c_2=0.7$ we demonstrate (see Fig.~\ref{Fig43}) that a bounce is possible at $U(\sigma)>0$ and the corresponding potential has a minimum.

 Numerical calculations show that in this case the conditions
$V_c<0$ and $4V_c-\sigma V_c^\prime>0,$ require $\sigma\in(0.230742,0.285248)$.

\section{The induced gravity models}
On using (\ref{poten}) and (\ref{connection_ind}), we construct the induced gravity model
\begin{equation}
S=\int dx^4\sqrt{-g}\left(U_{ind}R
-\frac12g^{\mu\nu}\tilde{\sigma}_{,\mu}\tilde{\sigma}_{,\nu}+V_{ind}\right),
\label{Sind}
\end{equation}
with the following potential
\begin{equation*}
\fl
V_{ind}=\frac{\gamma^2\sigma^4}{4^{\frac{1}{\beta}}\left(\sigma_0\sigma\right)^{6\Gamma(1-\beta)}U_0^2}\left[
c_1\left(\sigma^{6\beta\Gamma}+\sigma_0^{6\beta\Gamma}\right)^{\frac{2(1-\beta)}{\beta}}+c_2\left(\sigma^{6\beta\Gamma} -\sigma_0^{6\beta\Gamma}\right)^{\frac{2(1-\beta)}{\beta}}\right],
\end{equation*}
where $\Gamma\equiv \sqrt{\frac{1+6\gamma}{6\gamma}}$.

In order to obtain the general solution we use formulae (\ref{Na}) and (\ref{connection_ind}) and write the general solution in terms of the variables $\xi$ and $\eta$:
\begin{eqnarray}\label{asigmaXY}
   \sigma_{ind}&=&\sigma_0\left(\frac{\xi+\eta}{\xi-\eta}\right)^{1/(6\beta\Gamma)}, \\
  a_{ind}&=&\frac{\sqrt{U_0}}{\sigma_0}\sqrt{\frac{2}{\gamma}}\left(\frac{\xi-\eta}{\xi+\eta}\right)^{{1/(6\beta\Gamma)}}\left(\xi^2-\eta^2\right)^{1/(6\beta)},  \\
  N_{ind}&=&\frac{4U_0^{3/2}}{3\beta^2\sigma_0}\sqrt{\frac{2}{\gamma}}\left(\frac{\xi-\eta}{\xi+\eta}\right)^{{1/(6\beta\Gamma)}}\left(\xi^2-\eta^2\right)^{-1+1/(2\beta)}\,.
\end{eqnarray}

Thus, in order to obtain the general solutions for the induced gravity model (\ref{Sind}), one has to solve the same equations (\ref{equxieta})  as for the model with a minimally coupled scalar field and the potential (\ref{Wgeneral}). Note that for $\beta=1/3$ this induced gravity model is integrable for nonzero values of $K$ as well.

\section{Concluding remark}
On using the connections between exact solutions of the cosmological models with minimally coupled and non-minimally coupled scalar fields, one can obtain some interesting results.
We have applied this connection to show that the integrable cosmological model~\cite{Boisseau:2015hqa} with a constant $R$ belongs to a one-parameter set of integrable models with the same function $U_c$ and different potentials $V_c$. The corresponding integrable induced gravity models were also constructed.

A cosmological model, with a particular potential depending on hyperbolic functions, was intensively studied in a series of papers \cite{Bars:2010zh,Bars:2011th,Bars:2011mh,Bars:2012mt,Bars:2011aa} having a special emphasis on questions such as the possibility of crossing the Big Bang-Big Crunch singularity and an antigravity regime. Such considerations led to some discussions~\cite{Starobinsky1981,Sami:2012uh,Carrasco:2013hua,Bars:2013qna}. Remarkably, this model  has a conformally coupled counterpart in an  integrable model with a rather simple cosmological dynamics and a bounce solution, thus avoiding the cosmological singularity \cite{Boisseau:2015hqa,Boisseau:2015cda}.

On the other hand, this model with its potential, which includes fourth degrees of hyperbolic functions,  can be considered as a particular case of a wider class of models (case 7 in paper \cite{Fre}), for which, in the present paper,  we have constructed their conformally coupled and induced gravity analogues.
We note that sometimes the model with non-minimal coupling can be integrated on using some symmetry properties (see, for example~\cite{Borowiec:2014wva}), so it may be useful to consider conformally connected models both with minimal and non-minimal couplings in order to obtain new integrable models. In this paper we find integrable cosmological models with non-minimally coupled scalar fields with both open and closed FLRW metrics.

Let us emphasize that the models with a minimal coupling cannot have bouncing solutions since the second time derivative of the scale factor is proportional to the time derivative squared of the scalar field taken with the opposite sign.
Thus, the dynamics in models with a minimal coupling and that in their non-minimally coupled counterparts can be very different. We shall study this question in detail in a future publication \cite{future}. Finally, we wish to end by noting that the relation between the exact solutions in non-minimally coupled and minimally coupled cosmological models can be used to address some additional points in discussions concerning the crossing of the singularity~\cite{future}.

\textbf{Acknowledgments. }
The authors would  like to thank Alexey Toporensky for the useful discussions.
Research of E.P. and S.V. is supported in part by the RFBR grant 14-01-00707 and by the Russian Ministry of Education and Science under grant NSh-3042.2014.2. A.K. was partially supported by the RFBR grant 14-02-00894.

\end{document}